\documentclass[12pt]{iopart}
\expandafter\let\csname equation*\endcsname\relax
\expandafter\let\csname endequation*\endcsname\relax 
\usepackage{amsmath,amssymb, color}
\usepackage{graphicx}

\usepackage{ulem}

\newcommand{\ba}{\begin{eqnarray}}
\newcommand{\ea}{\end{eqnarray}}
\newcommand{\be}{\begin{equation}}
\newcommand{\ee}{\end{equation}}

\begin{document}

\title[A hybrid witness for single-photon entanglement]{Probing wave function collapse models with a classically driven mechanical oscillator}

\author{Melvyn Ho\footnotemark[3]\footnotetext{These authors contributed equally
to this work.}$^1$, Ambroise Lafont\footnotemark[3]$^1$, Nicolas Sangouard$^1$,\\ Pavel Sekatski$^2$}

\address{$^1$Department of Physics, University of Basel, Klingelbergstrasse 82, 4056 Basel, Switzerland\\
$^2$Institut for Theoretische Physik, Universitat of Innsbruck, Technikerstra{\ss}e 25, A-6020 Innsbruck, Austria\\
}

\begin{abstract}
We show that the interaction of a pulsed laser light with a mechanical oscillator through the radiation pressure results in an opto-mechanical entangled state in which the photon number is correlated with the oscillator position. Interestingly, the mechanical oscillator can be delocalized over a large range of positions when driven by an intense laser light. This provides a simple yet sensitive method to probe hypothetical post-quantum theories including an explicit wave function collapse model, like the Diosi \& Penrose model. We propose an entanglement witness to reveal the quantum nature of this opto-mechanical state as well as an optical technique to record the decoherence of the mechanical oscillator. We also report on a detailed feasibility study giving the experimental challenges that need to be overcome in order to confirm or rule out predictions from explicit wave function collapse models. 
\end{abstract}

\maketitle

\section{Introduction}
Post-quantum theories have been proposed which provide explicit wave function collapse models to explain how the classical world emerges from the quantum domain, see e.g. \cite{Ellis84, GRW86, GRW90, Gisin89, Diosi89, Penrose96}. Although the physics behind each collapse mechanism differs, they all operate as a spatial localization preventing massive objects to be in a quantum superposition of two or more positions. A possible approach to test them is to manipulate the motion of a mechanical oscillator through the radiation pressure. In this framework, it has been recently proposed to test collapse models by simply looking at the spectrum of the light driving the oscillator \cite{Bahrami14, Nimmrichter14} or through a spontaneous increase of the equilibrium temperature \cite{Diosi15}. Alternatively, we can look for a method to push the mechanical oscillator down to the quantum regime where its spatial position is largely delocalized and a technique to record the decay of spatial quantum coherences. Deviation from standard decoherence that occurs through the interaction with the environment \cite{Zurek03} might make it possible to confirm or rule out predictions from these hypothetic wave function collapse models. The proposals of Refs \cite{Bose97, Bose99, Marshall03} have followed this approach. They consist of first preparing quantum light, entangling it with the mechanical oscillator position and subsequently observing the oscillator decoherence through the decay of quantum properties of light. While recent proposals have shown how to relax some of the constraints on the opto-mechanical coupling strength, they still need non-classical light to start with \cite{Pepper12, Kleckner08, Sekatski14, Ghobadi14}. In the resolved-sideband regime, techniques benefiting from an optomechanical squeezing interaction \cite{Hofer11, Palomaki13} or based on conditioning \cite{Basiri-Esfahani12, Komar13, Vanner13a, Galland14} have been put forward to create quantum optomechanical states while avoiding the initial preparation of non-classical light.\\

In this work, we use laser light to drive a mechanical oscillator in the pulsed regime where the light pulse duration is much shorter than the mechanical period. In this regime, the mechanical oscillator can be cooled and manipulated without sideband resolution \cite{Vanner11, Wang11, Machnes12} as shown in an recent experiment \cite{Vanner13b}. The basic principle for manipulation is that the kick imparted by the light is proportional to the photon number. In particular, we show that when the oscillator is driven by an intense laser pulse where the photon number is inherently and largely undefined, this results in an opto-mechanical entangled state in which the oscillator is delocalized over a large range of positions. We build up an entanglement witness that can be used to reveal the quantum nature of this opto-mechanical state. We also show how to disentangle the light and the oscillator while using the light to record the decay of the oscillator spatial coherences and ultimately, to probe hypothetic deviations from standard decoherence. We discuss the experimental feasibility of this test bench for wave function collapse models by studying the effects of various measurement inaccuracies and finite cooling efficiencies.\\

\section{Creating optomechanical entanglement}
Consider the optical and mechanical modes of an optomechanical cavity with bosonic operators $a$ and $m$ respectively. The corresponding Hamiltonian is given by $H=\hbar \omega_m m^\dag m - \hbar g_0 a^\dag a (m^\dag+m)$ where $\omega_m$ is the mechanical frequency and $g_0=\frac{\omega_c}{L}\sqrt{\frac{\hbar}{2M\omega_m}}$ is the optomechanical coupling, L being the cavity length, M the effective mass of the oscillator and $\omega_c$ the cavity frequency. Further consider the ideal case where the mechanical mode is initially prepared in its motional ground state $|0\rangle_M.$ When a $n$ photon Fock state $|n\rangle_A$ impinges upon the oscillator, they induce a displacement of the mechanical state whose amplitude is time dependent $e^{i \frac{g_0^2 n^2}{\omega_m^2} (\omega_m t - \sin(\omega_m t))} |\frac{g_0 n}{\omega_m}(1-e^{-i \omega_m t})\rangle_M$ $|n\rangle_A$ \cite{Bose97}. The result of the interaction with a laser pulse $|\alpha\rangle,$ (where we assume that $\alpha \in \mathbb{R}$ all along the paper), i.e. with a Poissonian distribution of number states can be obtained directly from this expression. In particular, in the pulsed regime where the interaction time $\tau$ satisfies $\omega_m \tau \ll 1,$ (cf below for the exact conditions) the optomechanical state reduces to  
\begin{equation}
\label{optomechanical_state}
e^{-\alpha^2/2}\sum_{n\geq 0} \frac{\alpha^n}{\sqrt{n!}} |n\rangle_A |i n g_0 \tau e^{-i \omega_m t} \rangle_M
\end{equation}
a time $t$ after the interaction.

\begin{figure}[htbp!]
\begin{center} \includegraphics[width=0.6\textwidth]{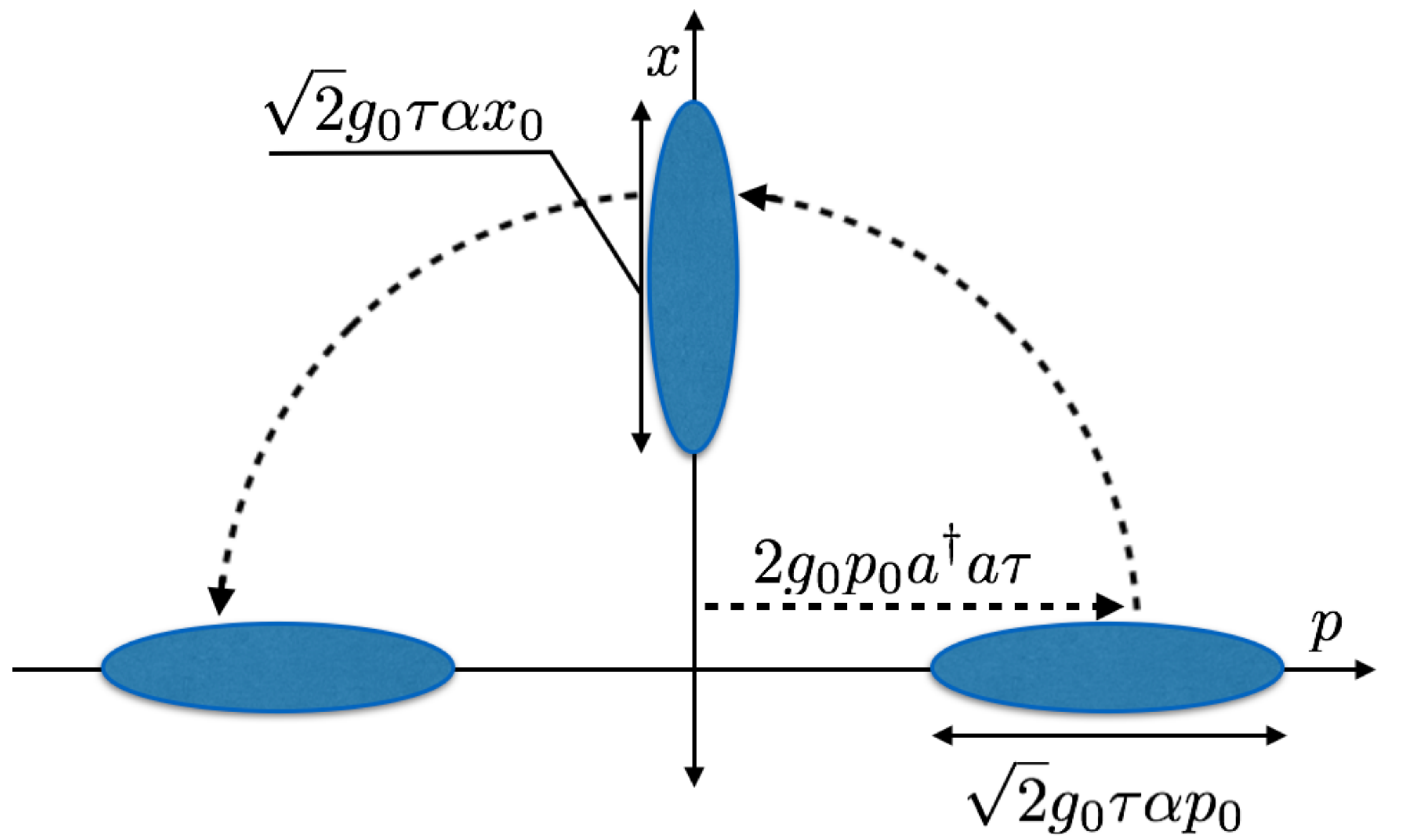} \end{center}
\caption{  Trajectory of the oscillator in phase space. With a short laser pulse, the mirror first experiences a momentum shift proportional to the respective photon number, and thereafter proceeds with oscillations in phase space. As the mirror travels, it assumes a large superposition of position states, and hence a significant position spread.}   
\label{Fig1}
\end{figure} 

The mechanical state involves coherent states $|i n g_0 \tau e^{-i \omega_m t} \rangle_M$ with amplitudes that depend on the photon number. 
Right after the interaction $(t=0),$ the photon number is entangled with the mechanical momentum. Then the mechanical oscillator rotates in phase space. After a quarter of a mechanical period $t=\pi /(2\omega_m),$ the photon number is entangled with the mechanical position before being entangled again with the mechanical momentum at $t=\pi /\omega_m$ and so on. Before we show how to reveal this optomechanical entanglement, let us note that the mechanical position is delocalized over the range $\langle \Delta X _M \rangle=x_0\sqrt{(1+2 g_0^2\alpha^2 \tau^2)}$ on average which can be made much larger than the mechanical zero-point fluctuation $x_0=\sqrt{\frac{\hbar}{2M \omega_m}}$ with an intense driving laser, even in the experimentally relevant regime where $g_0 \tau \ll 1$ $(g_0 \ll \omega_m \ll 1/\tau).$ As the characteristic timescale of the wave-function collapse models decreases with $ \Delta X _M $ some of them are expected to degrade the quantum properties of the state (\ref{optomechanical_state}) on timescales that can be accessed experimentally even when dealing with small effective masses, as we show below. \\

\section{Revealing optomechanical entanglement}
The question that we address in this paragraph is how to detect the quantum nature of the optomechanical state (\ref{optomechanical_state}). The entanglement witness that we propose is inspired by Ref. \cite{Tan99}. Concretely, for two pairs of local observables $( A_1,  A_2)$ and $( B_1,  B_2)$ and a separable state $\rho_{AB}$, the following inequality holds
$
\sqrt{\Delta^2( A_1-  B_1)\Delta^2( A_2-  B_2)} \geq \frac{1}{2}\left(| \langle [ A_1,  A_2] \rangle | + | \langle [ B_1,  B_2] \rangle |\right)
$
where $\Delta^2( A_1-  B_1) = \tr \rho_{AB} ( A_1-  B_1)^2 - \left(\tr \rho_{AB} ( A_1-  B_1)\right)^2$ stands for the variance. The idea is that the observables satisfy the Heisenberg uncertainty relation locally whereas the pairs of observables $(A_1, B_1)$ and $(A_2, B_2)$ can only be classically correlated for a separable state. The aim is thus to find two couples of observables $( A_1,  B_1)$ and $( A_2,  B_2)$ such that the variances $\Delta^2( A_i-  B_i)$ do not increase (decrease) during the evolution while the commutators $|\langle [ A_1,  A_2] \rangle|$ and $\langle [ B_1,  B_2] \rangle$ increase (stay constant). In our case, the optomechanical interaction shifts the oscillator momentum by the photon number (times the interaction strength) suggesting $ A_1= P_M $ and $ B_1=\sqrt{2} g_0 \tau a^\dag a.$ Similarly, the light field acquires a phase proportional to the position of the oscillator. While the phase $\varphi$ is not a physical observable, for a coherent state it can be indirectly accessed through homodyne detections   $\langle P_\ell\rangle = \sqrt{2} \alpha \sin \varphi.$ Therefore, we choose $ A_2 = \sqrt{2} \alpha \sin(\sqrt{2} g_0 \tau  X_M)$ and $ B_2 =  P_\ell.$ This leads to the following inequality 
\begin{eqnarray}
\nonumber
&&\sqrt{\Delta^2 ( P_M - \sqrt{2} g_0 \tau a^\dag a) \Delta^2(\sqrt{2} \alpha \sin(\sqrt{2} g_0 \tau  X_M) -  P_\ell)} \\
\label{ent_witness}
&&\geq \frac{g_0 \tau}{2} \left(  \sqrt{2} \alpha | \langle \cos ( \sqrt{2} g_0 \tau X_M)\rangle | + | \langle X_\ell \rangle |\right).
\end{eqnarray}
If the results of measurements do not fulfill this inequality, we can conclude that the light field and the mechanical oscillator are entangled. In particular, this inequality does not hold for the state (\ref{optomechanical_state}) if
\begin{equation}
\label{condition_violation}
\alpha^2 \geq \frac{1}{16 (g_0 \tau)^2}.
\end{equation}
This condition is obtained under the assumption $g_0 \tau \ll 1.$ It can be understood as a requirement on the laser intensity to significantly enlarge the mechanical zero point spread -- a condition that is necessary to correlate the photon number and the mechanical momentum. \\

\begin{figure}
\begin{center} \includegraphics[width=0.6\textwidth]{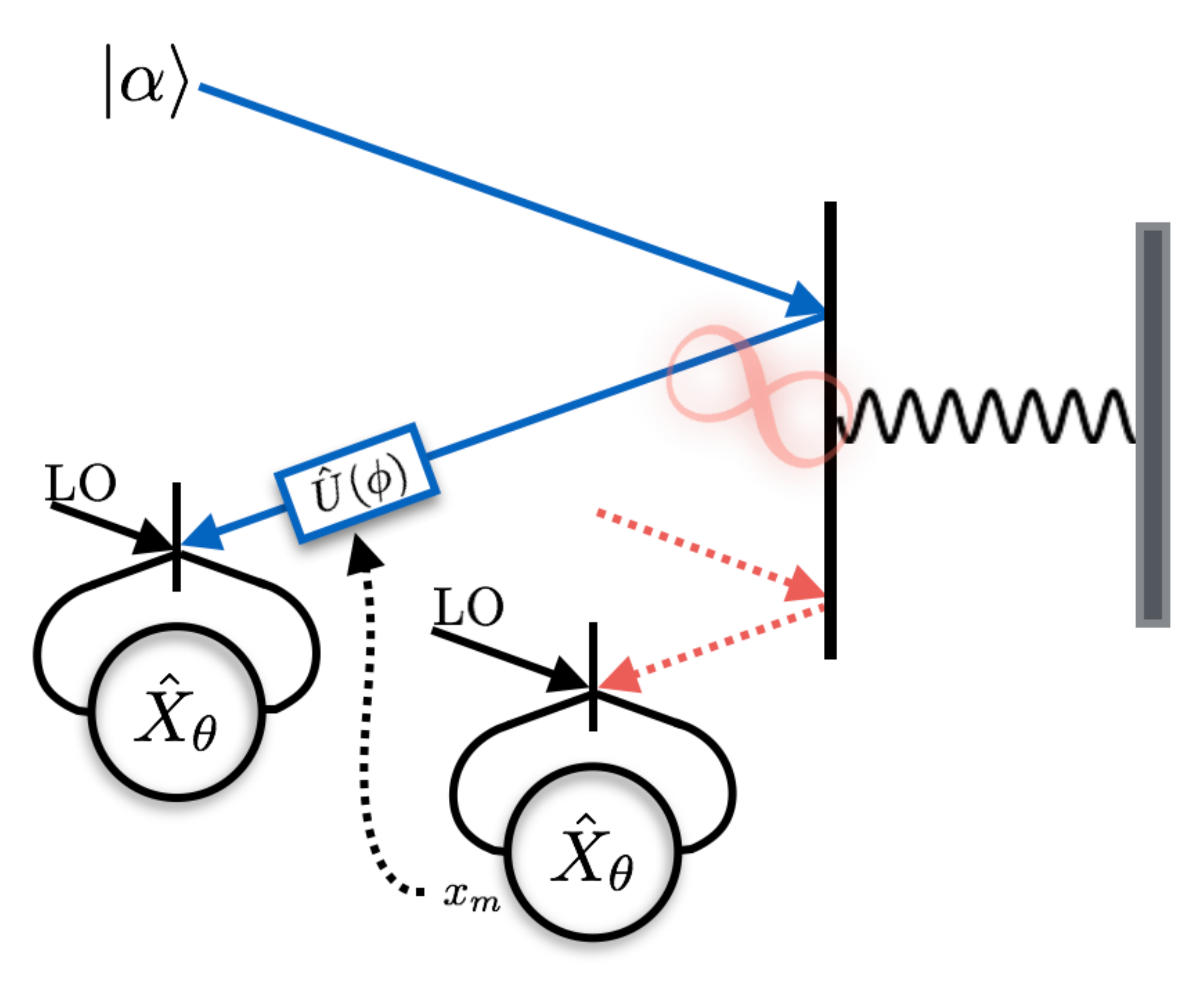} \end{center}
\caption{Once the light pulse $|\alpha\rangle$ interacts with the mechanical device, the photon number is entangled with the mechanical momentum. To detect this optomechanical entangled state, we need to look at the correlations between i) the photon number and the mechanical momentum, ii) the quadrature of the light and the mechanical position. The latter can be performed by using the fact that the phase shift in the reflected light is proportional to the position of the mirror. To record the mechanical decoherence, the mechanical oscillator is first disentangled from the light by measuring the mechanical position at a time which is a multiple of half the mechanical period (See Fig. \ref{Fig1}). A feedback loop is then needed to correct the phase of the light depending on the result of the measurement of the mechanical position. The oscillator decoherence translates into a phase noise on the light that can be observed with an homodyne detection. Deviation from environmental decoherence might make it possible to confirm or rule out predictions form explicit wave function collapse models.} 
\label{Fig2}
\end{figure}

\section{Recording the mechanical oscillator decoherence}
We first explain how to proceed with recording the decoherence of the mechanical oscillator:

\begin{itemize}
\item[a)] A short pulse of duration $\tau$ generates the entangled optomechanical state  (\ref{optomechanical_state}).
\item[b)] After a variable delay $t = k \frac{\pi}{\omega_m}$, another short pulse is used to measure the mechanical position, where no information is revealed about the photon number.   
\item[c)] The phase of the light pulse in (a) is shifted using the measurement outcome of step (b).
\item[d)] The phase quadrature of the outgoing light is measured.
\end{itemize} 

%c.f. Fig. \ref{Fig1}.       a) Firstly, as in (\ref{optomechanical_state}), a short pulse of duration tau generates an entangled optomechanical state, b) After a variable delay $t = k\frac{\pi}{\omega_m}$ another short pulse is used to measure the mechanical position, where no information is revealed about the photon number. c) the phase of the light pulse in step (a) is shifted depending on the measurement outcome of step (b) and its phase quadrature is measured. }  

In the ideal case without decoherence, the conditional state of the light $|\alpha e^{i(-1)^k\sqrt{2} g_0 \tau x_m} \rangle$ has a phase which depends on the result $x_m$ of the measurement of the oscillator position. Correcting the phase of the light with a feedback loop yields a photonic state that is identical to the initial light state $|\alpha \rangle$.

Things are different if we take the oscillator decoherence into account.  These decoherence models, conventional or otherwise, operate as a decay of spatial coherences 
$
\dot{|x\rangle\langle y|}=- \gamma (x-y) |x\rangle\langle y|,
$
where $|x\rangle$ and $|y\rangle$ are position eigenstates. The exact expression of $\gamma$ will differ according to the decoherence model being described \cite{footnote_gamma}, but all operate as a spatial localisation as a function of the distance $|x-y|$, resulting in a phase noise on the light once the position of the mechanical device is measured \cite{Sekatski14}. Instead of  $|\alpha\rangle$, in the presence of decoherence, one obtains (see \cite{Sekatski14}, Supplemental Material)

\begin{equation}
\rho=\int d\varphi \-\ \tilde{\xi}(\varphi) \-\ |\alpha e^{i \varphi} \rangle\langle \alpha e^{i\varphi} |
\end{equation}
where $\tilde{\xi}(\varphi) = \frac{1}{2\pi} \int_{-\infty} ^\infty  d\eta \  e^{i\eta\varphi} e^{-k \frac{\pi}{\omega_m} \langle \gamma(2 \eta g_0 \tau x_0 \sin(\theta))\rangle},$ with 
$\langle \gamma(X \sin (\theta))\rangle=\frac{1}{\pi} \int_0^{\pi} \gamma(X \sin(\theta)) d\theta$ indicating an average over half a period of the mechanical oscillator. The problem of identifying the oscillator decoherence is thus equivalent to characterizing a phase noise channel with light -- this characterization is more accurate when probed with large photon numbers. For example, by measuring the quadrature $X_\ell^{\theta}$ of the light field, we get

\begin{eqnarray}
\label{moment_X_1}
\langle  X_\ell^{\theta} \rangle &=&\int_{-\infty} ^{\infty} d \varphi \tilde{\xi}(\varphi) \langle \alpha e^{i \varphi}|\frac{a e^{- i \theta} + a^\dag e^{i \theta}}{\sqrt{2}}  |\alpha e^{i \varphi}\rangle \nonumber \\
&=&\int_{-\infty}^{\infty} d \eta \ e^{-k \frac{\pi}{\omega_m} \langle \gamma(2 \eta g_0 \tau x_0 \sin(\theta))\rangle} \nonumber  \\
&&\frac{1}{\sqrt{2}}\left(\frac{1}{2\pi} \int_{-\infty} ^{\infty}  \ d  \varphi \  e^{i (\eta+1) \varphi} \alpha e^{-i\theta} + \frac{1}{2\pi} \int_{-\infty} ^{\infty} \  d \varphi \  e^{i (\eta-1) \varphi} \alpha e^{+i\theta} \right)  \nonumber \\
&=& \sqrt{2} \alpha \xi(1) \cos \theta,
\end{eqnarray} 

and similarly 
\begin{eqnarray}
\label{moment_X_2}
&& \langle  (X_\ell^{\theta})^2 \rangle = \frac{1}{2} + \alpha^2 \left(1+\cos \left(2\theta\right) \xi(2)\right),
\end{eqnarray}
where $\xi$ is defined by $\tilde{\xi}(\varphi) = \frac{1}{2\pi} \int d\eta e^{i\eta\varphi} \xi(\eta)$ i.e. 
\begin{equation}
\xi(\eta)=e^{-k \frac{\pi}{\omega_m} \langle \gamma(2 \eta g_0 \tau x_0 \sin(\theta))\rangle}.
\end{equation}
Immediately after the optomechanical interaction (k=0), the mechanical oscillator has not yet undergone any decoherence and $\xi(\eta)=1$ $\forall \eta.$ The phase of the light is well defined and the variance of $|\alpha\rangle$ is $1/2.$ In the limit where $k \rightarrow \infty,$ the spatial coherences of the mechanical oscillator vanish $\xi(\eta)=0$ $\forall \eta$. The light field is described by a mixture of coherent states with random phases and its spread in the phase space tends to $1/2+\alpha^2.$ Focusing on $ X_\ell^{\theta=\pi/2}= P_\ell,$ the time taken to double the distribution of possible results, for example, is given by 
\begin{equation}
\frac{1}{2\alpha^2 \langle \gamma(4 g_0 \tau x_0 \sin(\theta))\rangle}
\end{equation}
at the leading order in $\alpha$. We see that the use of intense laser pulses reduces the time it takes to observe the effect of collapse models. As particular examples, we compare conventional decoherence with explicit collapse models as proposed by Ellis and co-workers \cite{Ellis84} on the one hand and Diosi \& Penrose \cite{Diosi89, Penrose96} on the other hand. We show that with a combination of large $\alpha$ and small thermal dissipation, we can find experimentally feasible parameters in which the standard decoherence time is longer than the timescale of these collapse models, hence opening the way to confirm or rule out their predictions. Note that for standard decoherence processes where $\gamma(x) = C x^2,$ $\xi(\eta)=e^{-k \frac{2\pi}{\omega_m}  C (g_0 \tau x_0)^2 \eta^2}$. In this case $\xi(2)/\xi(1)^4=1.$ For the Diosi \& Penrose model, $\xi(\eta)$ decreases less rapidly with $\eta$ such that the ratio $\xi(2)/\xi(1)^4$ can be much larger than one. As this ratio can be accessed through the two first moments of $X_\ell^{\theta}$ \eqref{moment_X_1}-\eqref{moment_X_2}, this provides an unambiguous criteria to distinguish standard decoherence with the Dioisi \& Penrose model. This is particularly relevant in the regime where $g_0 \tau x_0$ is comparable to the parameter that is used to define the mass distribution in the Diosi \& Penrose model.\\

\section{Experimental constraints}
We now address the question of the feasibility in detail. First, our results have been derived under the assumption that the dynamics falls within the  pulsed regime. This requires
\begin{equation}
\label{condition_linear}
\alpha^2 (g_0 \tau)^2 \omega_m \tau \ll 6
\end{equation}
which provides an upper bound on the photon number given the optomechanical coupling, the pulse duration and the mechanical frequency placing an upper limit on the extent to which we can violate (\ref{ent_witness}).

%In particular if $\alpha^2 (g_0 \tau)^2 \omega_m \tau = 0.6,$ we check that the inequality (\ref{ent_witness}) is significantly violated.

Let us now consider various imperfections. First, let the measurement of the oscillator position be subject to a Gaussian noise with a spread $\delta x.$ To reveal entanglement, one needs $\frac{\delta x^2}{x_0^2} \sim 1.5$ in the limit $(4g_0 \tau \alpha)^2 \gg 1,$ i.e. to resolve the zero point spread. The requirement is more demanding for recording decoherence. If the mechanical position is not precisely measured, the phase of the light cannot be accurately corrected leading to phase noise. This effect has to be smaller than the phase noise induced by decoherence. Since we are interested in the regime where decoherence doubles the variance of $P_\ell,$ we need

\begin{equation}
\label{accuracy_position_deco}
\frac{\delta x^2}{x_0^2} \leq \frac{1}{4( g_0 \tau \alpha)^2}.
\end{equation} 
Note that the mechanical position can be measured by homodyning a light pulse that has been sent into the mechanics, since the phase of the reflected light pulse depends on the mechanical position. It has been shown in Ref. \cite{Vanner11} that for an input drive with duration $\ln 2/\kappa,$ the achievable precision $\delta x$ depends on the photon number $N_p$ through the formula $\delta x = x_0 \frac{\kappa}{\sqrt{5} g_0 \sqrt{N_p}}.$ The limitation for $N_p$ and hence on the precision is given by the power $P_p$ that can be used before heating significantly the surrounding bath and ultimately by the power that can be homodyned before photo-detection saturates. 

Similarly to the requirement on the precision of the measurement of the oscillator position, the measurement of the quadrature of the light field needs to be accurate. Consider an imperfect quadrature measurement  in which the phase of the local oscillator follows a Gaussian distribution with a standard deviation $\sigma.$ This is equivalent to a phase noise on the light field and a perfect local oscillator. We found that the proposed witness can reveal optomechanical entanglement if 
\begin{equation}
\label{accuracy_homodyne_ent}
\sigma \leq 2 g_0 \tau.
\end{equation}
To accurately record the decoherence of the mechanics, we find 
$
\sigma \leq \frac{1}{2\alpha}.
$
In the limit $g_0 \tau\alpha \gg 1$ (\ref{condition_violation}), the constraint for observing entanglement is the most demanding.

If the mechanical oscillator is not initially in its ground state, but in a thermal state with a mean occupation $n_{\text{th}},$ the variance of the light is unchanged but the correlations between the light and mechanics decreases. To detect entanglement, one needs   
\begin{equation}
\label{cond_occupation}
n_{\text{th}}+\frac{1}{2} \leq 8 (g_0 \tau \alpha)^2 
\end{equation} 
in the limit where $n_{\text{th}} g_0 \tau \ll 1.$ Various cooling schemes have been proposed in the pulsed regime \cite{Vanner11,Machnes12}. For example, it has been shown in Ref. \cite{Vanner11} that measuring the mechanical position with two pulses (of duration $\frac{\ln 2}{\kappa}$ and containing each $N_p$ photons) delayed by a quarter of a mechanical period reduces the mechanical excitation to an effective thermal occupation 
$
n_{\text{eff}}= \frac{1}{2} \left(\sqrt{1+\frac{\kappa^4}{g_0^4 \bar N_p^2}}-1\right).
$
For $g_0 \sqrt{N_p} > \kappa,$ this results in $n_{\text{eff}} \ll 1,$ i.e. ground state cooling. Again, the limitation on $\bar N_p$ is the power $\bar P_p$ than can be homodyned and that can be undergone by the mechanics without increasing the temperature of the environment. 

To summarize, let us consider a given optomechanical system with fixed mechanical and cavity frequencies, effective mass and cavity length, i.e. $g_0$ and $\omega_m$ are given. We choose a pulse duration $\tau$ as large as possible to relax the constraint on the precision of the homodyne detection (\ref{accuracy_homodyne_ent}). In particular, for $\tau \sim \frac{\ln{2}}{\kappa},$ the maximum value is set by the cavity finesse. We then choose the largest possible $\alpha$ so that the dynamics still holds in the pulsed regime (\ref{condition_linear}). We found that $\alpha^2=0.6/((g_0 \tau)^2 \omega_m \tau)$ provides a significant violation of the inequality (\ref{ent_witness}). The detection of entanglement then gives a constraint on the initial effective occupation number of the optomechanical system through the formula (\ref{cond_occupation}). Lastly, the formulas (\ref{accuracy_position_deco}) and (\ref{accuracy_homodyne_ent}) give the requirement on the measurement precision. 
The temperature of the environment and the mechanical quality factor are such that  standard decoherence should operate on times scales longer than the collapse models we wish to test. 
\\
 \section{Experimental feasibility}
For concreteness, we focus on a mechanical oscillator with an effective mass $M=60$ ng and a frequency $\omega_m = 2\pi \times 20 \times 10^3$ s$^{-1}$ that is used as one of the mirrors of a Perot-Fabry cavity with length 0.5 cm, resulting in an optomechanical coupling $g_0 \sim 2\pi \times 100$ s$^{-1}.$ We consider a cavity finesse of $1.5 \times 10^5$ which corresponds to the finesse of the Perot-Fabry cavity implemented with micromirrors in Ref. \cite{Muller10}. This leads to light pulses with a duration $\tau \sim 1.1$ $\mu$s with a mean photon number of up to $\alpha^2 \sim 8.6 \times 10^{6}$ i.e. a power is 1 $\mu$W. 
To reveal entanglement, the thermal occupation number of the mechanics has to satisfy $n_{\text{th}} \leq 34$ which can be achieved with a power $\bar P_p \sim 1.6$ nW and the phase of the local oscillator that is used for homodyning the light has to be set with an accuracy of $\sigma \sim 0.1$ $\deg.$ This also takes a base temperature of $T \sim 400$ mK and a mechanical Qm factor of $\sim 10^6.$  To record the decoherence of the mechanics, the mechanical position needs to be resolved with an accuracy $\frac{\delta x}{x_0} \leq 0.24$  which can be achieved with a power $P_p$ of  0.38 $\mu$W. Furthermore, we found that the model by Ellis and co-workers takes about $5 \times10^{-5}$s to double the spread of results of $P_\ell$, which would be testable with the proposed device with Qm $\sim 1.5\times 10^7$ and $T\sim 20$ mK. The model of Diosi \& Penrose might also be testable despite the known ambiguity with respect to the mass distributions. Under the assumptions that the mass is distributed over spheres corresponding to the size of the atomic nuclei, it takes about $2\times 10^{-8}$s to double the spread of results of $P_\ell$. This should be testable with Qm $\sim 10^5$ and $T \sim 300$ mK.\\

\section*{Conclusion}
We have shown how a simple laser light driving a mechanical oscillator in the pulsed regime results in an optomechanical entangled state. We have shown how to reveal the quantum nature of this state and how to use it to probe the oscillator decoherence. While we have focused on a realization with the trampoline resonator envisioned in Ref. \cite{Kleckner11}, various systems might be used to implement our proposal and ultimately, to test explicit wave function collapse models. We found for example that despite its small mass, the zipper cavities of Ref. \cite{Cohen13, Eichenfield09} might be used to test the model of Diosi \& Penrose model with a bath temperature of $T \sim 200$ mK if their mechanical Q factor is pushed to $Q \sim 10^7.$ \\

\section{Acknowledgements} We thank F. Frowis and E. Oudot for many discussions. This work was supported by the National Swiss Science Foundation (SNSF), Grant number PP00P2-150579 and the Austrian Science Fund (FWF), Grant number J3462 and P24273-N16.\\

\section*{Appendix} Here, we clarify the constraint for the system to be in the pulsed regime.

Immediately after the interaction, the system is in the state 

\begin{eqnarray}
|\psi\rangle = e^{i \lambda {(a^\dag a)}^2}  D_M( a^\dag a \  \beta )  D_\ell(\alpha) |0\rangle_{\ell} |0 \rangle_M  \nonumber,
\end{eqnarray}
where $\beta = \frac{g_0}{\omega_m}(1-e^{- i \omega_m \tau})$ and $\lambda = \frac{(g_0 \tau)^2(\omega_m \tau)}{6} + \mathcal{O}((\omega_m \tau)^2)$. $D_{M(\ell)}$ refers to the displacement operator on the mirror (light) system respectively.

Performing a phase shift $e^{-2 i \alpha^2 \lambda 
a^\dag a}$ and taking the overlap of this state to the ideal case, we find

\begin{eqnarray}
&&|\langle 00| \left[ D^\dag_{\ell}(\alpha) D^\dag_{M}(a^\dag a \beta) \right] \left[ e^{-2 i \alpha^2 \lambda 
a^\dag a } e^{i \lambda (a^\dag a)^2} D_{M}(a^\dag a \beta)D_{\ell}(\alpha) \right]|00 \rangle|^2  \nonumber \\
&=&|\langle 00 | D_l^\dag (\alpha) e^{-2 i \alpha^2 \lambda 
a^\dag a + i \lambda (a^\dag a)^2} D_l (\alpha)|00\rangle|^2 \nonumber  \\
&=&|\langle 00 | e^{  + i \lambda [(a^\dag a)^2 + \sqrt{2} \alpha \{ a^\dag a, X_{\ell} \} + 2\alpha^2 X_{\ell} ^2] } |00\rangle|^2  \nonumber \\
&\approx&|1+ i \lambda \alpha^2|^2 \nonumber .
\end{eqnarray}

Here we use the fact that $D_l ^\dag(\alpha)  (a^\dag a)  D_l(\alpha) = a^\dag a + \sqrt{2} \alpha X_{\ell} + \alpha^2$, and requiring that $\lambda \alpha^2 \ll 1$ yields the requirement that $\alpha^2 (g_0 \tau)^2 \omega_m \tau \ll 6$.

The fact that the light pulse is not sufficiently short, can lead to a mirror with a different starting position in phase space, even right after the mirror-light interaction time. This can be seen from the displacement of the mirror state.

\begin{eqnarray}
D_M(a^\dag a \beta) = D_M \left[ 2\frac{g_0}{\omega_m} a^\dag a \left( \sin^2(\frac{\omega_m \tau}{2}) + i \sin (\frac{\omega_m \tau}{2}) \cos (\frac{\omega_m \tau}{2}) \right) \right] \nonumber 
\end{eqnarray}

To disentangle the light from the mechanics then, the mirror measurements should be adjusted by $\frac{\omega_m \tau}{2}$.


\begin{thebibliography}{10}
\bibitem{footnote_gamma}  Note that for environmental induced decoherence $\gamma(|x-y|) = \frac{M k_b T \omega_m }{\hbar^2 Q_m} |x-y| ^2 $. For the model proposed by Ellis \cite{Ellis84}, $\gamma(|x-y|) \approx \frac{c^4 M^2 m_0^4  }{\hbar^3 m_p^3} |x-y| ^2 $. For the model of Diosi \& Penrose \cite{Diosi89, Penrose96}, $\gamma(|x-y|) = \frac{8 \pi G M m_n }{\hbar} (\frac{6}{5a} - \frac{1}{|x-y|})$ for $|x-y| > 2a$, and $ \frac{8 \pi G M m_n }{\hbar} (\frac{6}{5a} -   \frac{12a^2 - 5 |x-y| ^2}{10a^3} + \frac{|x-y| ^5 - 30a^2 |x-y| ^3}{160a^6})$ otherwise.
\end{thebibliography}
\end{document}